\documentclass[aps,prl,reprint,showpacs,showkeys,longbibliography,superscriptaddress,amsmath,amssymb,floatfix,footinbib,a4paper]{revtex4-2}

\usepackage{graphicx}
\usepackage[usenames,dvipsnames]{color}
\usepackage{xcolor}
\usepackage{epsfig}
\usepackage{wasysym}
\usepackage{natbib}
\usepackage{bm}
\usepackage[normalem]{ulem}
\usepackage{siunitx}

\newcommand{\br}{{\bf r}}

\newcommand{\bj}{{\bf j}}
\newcommand{\be}{{\bf e}}
\newcommand{\bff}{{\bf f}}

\newcommand{\bV}{{\bf V}}

\newcommand{\bG}{{\bf G}}
\newcommand{\bK}{{\bf K}}

\newcommand{\bOmega}{{\boldsymbol{\Omega}}}

\newcommand{\meanD}{{\Delta}}

\newcommand{\eq}[1]{Eq.~(\ref{#1})}
\newcommand{\eqs}[1]{Eqs.~(\ref{#1})}

\newcommand{\eref}[1]{(\ref{#1})}

\newcommand{\rcite}[1]{Ref.~\cite{#1}}

\begin{document}

\title{Ionic self-phoresis maps onto correlation--induced self-phoresis}

\author{Alvaro Dom{\'\i}nguez}
\email{\texttt{dominguez@us.es}}
\affiliation{
  F{\'i}sica Te{\'o}rica, Universidad de Sevilla, Apdo.~1065, 
41080 Sevilla, Spain}
\affiliation{Instituto Carlos I de F{\'i}sica Te{\'o}rica y Computacional,
  18071 Granada, Spain}

\author{Mihail N.~Popescu}
\email{\texttt{mpopescu@us.es}}
\affiliation{
  F{\'i}sica Te{\'o}rica, Universidad de Sevilla, Apdo.~1065, 
41080 Sevilla, Spain}

\date{\today}

\begin{abstract}
We re-examine the self-phoresis of a particle that releases(removes) pairs of ions into(from) the electrolyte solution. We show analytically that in the linear regime 
the mathematical description of this system maps  onto that of the correlation--induced (self-)chemophoresis (CICP). This connection provides a unifying perspective of the two phenomena, within which one recovers and extends recent predictions as particular
instances of CICP.  Conversely, ion-phoretic particles are identified
as candidates for experimental investigations into the rich variety of
motility patterns predicted by CICP.
\end{abstract}

\maketitle


The emergence of self-motility for chemically active colloidal
particles suspended in an aqueous solution of their ``fuel'' (e.g.,
hydrogen peroxide for certain bi-metallic rods
\cite{Paxton2004,Fournier-Bidoz2005} or platinum-capped dielectric
particles \cite{Howse2007,Baraban2012}) has been the topic of many
experimental and theoretical investigations since the first
experimental reports were published (insightful reviews of these
developments can be found in, e.g., Refs.
\cite{Ebbens2010,BDLR16,DEY2016,Palacci2017,Moran2017}). In many
instances, the motility was successfully addressed as self-phoresis
(e.g., self-chemophoresis
\cite{Golestanian2005,Koplik2013,Palacci2013,Simmchen2016},
self-thermophoresis through a single-component fluid \cite{Kroy2016},
self-chemophoresis via demixing of a critical binary liquid mixture
\cite{Volpe2011,Wurger2015,samin2015self}, or self-electrophoresis
\cite{Moran2011,Seifert2012,Wang2016,Liverpool2017,Tasinkevych2019}).
That is, motility was interpreted to emerge as a \textit{classic}
phoretic linear response to self-generated, rather than externally
imposed, non-equilibrium inhomogeneities (in the chemical composition or
in the temperature of the solution)
\cite{Anderson1989,Golestanian2005,Kapral2007,Michelin2014,Moran2017}. Recently,
the standard scenario of chemophoresis has been investigated beyond
the usual ideal-gas approximation for the solute by accounting for
thermodynamic correlations \cite{DPRD20,DoPo22}. The result is the
identification of a novel mechanism of phoretic motility, the
so-called \emph{correlation--induced chemophoresis} (CICP), which provides a
paradigm--breaking example that self-phoresis involves in general a
non-equilibrium response factor, so that an interpretation as phoresis
in self-generated gradients is not possible \cite{DPRD20}.

In an ubiquitous experimental setup, the particles are charged and the
chemical activity consists in the release of ionic radicals in the
ambient solution
\cite{Paxton2004,Moran2011,Wang2016,Bayati2016,Brown2017,Corato2020,Moran2017,Asmolov2022}.
Additionally, there can be electrical currents through the solution
and the particle in the case of redox reactions at bi-metallic rods
\cite{Paxton2004,Moran2011}, or, as argued by
Refs.~\cite{Ebbens2014,BrPo14,Liverpool2017,Tasinkevych2019}, when a
dielectric sphere is covered by a layer of catalyst (a metal or
enzymes) of varying thickness.

Recent studies, complementing the earlier works by
Refs.~\cite{Bayati2016,Brown2017}, have reported a number of
interesting features for the case of a spherical particle, such as
motility of uncharged particles with non-uniform ionic activity
\cite{Corato2020} or of non-uniformly charged particles with uniform
activity \cite{Cruz2024}. In particular, in the former case the
phoretic velocity was noted to be quadratic in the activity
\cite{Corato2020}, which evades an interpretation as classic phoresis
under self-generated gradients.

Motivated by these insightful observations, here we re-examine a
simple model for ionic self-phoresis of a dielectric particle that
releases (or removes) pairs of ions in the electrolyte solution.  The
theoretical analysis is impaired by the nonlinear coupling between the
ionic distribution and the electric field
\cite{Moran2011,Bayati2016,Liverpool2017,Brown2017,Corato2020,Vinogradova2022,Cruz2024};
accordingly, the use of perturbative expansions has become a standard
approach
\cite{Prieve1984_JFM,Prieve1984_SPM,Prieve1987,Prieve1988,Dukhin1993}.
In the linear regime of small surface charge and activity of the
particle, we show the unexpected mapping of the mathematical
description of this model onto that of the recently reported CICP. On
the one hand, this result provides a conceptually clear, unifying
perspective which blurs the conceptual distinction between ionic self-phoresis and neutral self-chemophoresis: recent predictions of motility for certain activity and charge patterns then emerge, and are extended, as particular instances of the
``selection rules'' derived in the CICP model. On the other hand, it
highlights ionic self-phoresis as a promising option for experimental
investigations into the rich variety of patterns of motility predicted
by CICP (see, e.g., the ``phase diagram'' introduced in
Ref.~\cite{DoPo22}).

\noindent\textit{Model system.--}
The model we consider consists of a rigid and impermeable dielectric
colloidal particle, immersed in a liquid electrolyte solution which is
kept at a constant temperature $T$. For simplicity, we consider the
case of a spherical shape (radius $R$) for the particle, of a
symmetric electrolyte (thus two species of ions carrying charges
$\pm q$, $q>0$, respectively), and of the dielectric permittivity
$\epsilon$ of the electrolyte solution being much larger than that of
the particle. The concentration of the two ionic species are equal to
$c_0 \neq 0$ far from the particle. The particle carries a surface
charge $\sigma_s \, \mathbb{S}(\br_p)$, where $\sigma_s$ is a
characteristic value of the surface charge (e.g., its maximum over the
surface) and the dimensionless surface field $\mathbb{S}(\br_p)$,
where $\br_p$ denotes any point on the surface of the particle,
describes the distribution of charge over the surface. The activity of
the particle is characterized by the rate per unit surface
$\mathcal{A} \,\mathbb{A}(\br_p)$ at which pairs of ions are released
into ($\mathbb{A}>0$) or removed from ($\mathbb{A}<0$) the solution at
the position $\br_p$; here, $\mathcal{A}>0$ is the maximum absolute
value over the surface and the dimensionless surface field
$\mathbb{A}(\br_p)$ describes the pattern of activity \footnote{We
  have chosen the case that the ions taking part in the catalytic
  reaction are of the same species as the ones already present in the
  background electrolyte for reasons of conceptual simplicity. The
  study can be straightforwardly extended to the case of different ion
  pairs, but at the expense of significantly more cumbersome
  algebra.}. The ions diffuse in the solution with diffusion constants
$D_\pm$; the associated mobilities are given by the Stokes-Einstein
relation as $\Gamma_\pm = \beta D_\pm$, where $\beta = 1/(k_B
T)$. Finally, the system ``particle + electrolyte solution'' is
assumed to be in mechanical isolation, i.e., there are no external
forces or torques acting on either the particle or the fluid: this is
the characteristic feature of phoresis which sets it apart from other
transport phenomena \cite{Derjaguin,Anderson1989}, so that a
self-phoretic particle can be actually qualified as a ``swimmer''.

When the particle is inactive, the system is in an equilibrium state
(fixed by a distant heat bath and reservoirs of ions and solvent), in
which the particle and the fluid are motionless. Upon turning on the
activity, we assume that a non-equilibrium steady state is
established, in which the particle moves and hydrodynamic flow is
induced in the electrolyte. Motivated by the observations in typical
experimental realizations of active particles, we assume overdamped
motion of the particle, while the fluid flow occurs at small Reynolds
and Mach numbers (``creeping flow''). Additionally, the particle
motion can be assumed slow when compared with the diffusion of the
ionic species. Accordingly, the state of the solution is characterized
by the instantaneous incompressible flow of the solution and the
stationary concentrations $c_\pm(\br)$ of each ionic species at small
P{\'e}clet number (i.e., convection is neglected). An electric field
described by a potential $\psi(\br)$ will be also induced due to the 
local charge imbalances; owing to the assumed large contrast in dielectric
constants, the electric field is basically confined to the electrolyte
domain.

\noindent\textit{Steady--state distribution of ionic species.--}
The profiles $c_\pm(\br)$ follow from the conservation of ionic
species in the bulk expressed in terms of the ion density currents
$\bj_{\pm}$,
\begin{equation}
\label{eq:cons_ion}
\nabla \cdot \bj_{\pm} = 0\,,~~\bj_{\pm} = \Gamma_\pm \bff_\pm\,.
\end{equation}
The thermodynamic force densities $\bff_\pm$ are determined from the assumption of local
equilibrium, consistently with the assumption of slow particle dynamics, as
\begin{equation}
  \label{eq:f}
  \bff_\pm(\br)= - c_\pm(\br) \nabla\mu_\pm(\br),
\end{equation}
in terms of the local chemical potentials $\mu_\pm(\br)$. The latter are obtained as
\begin{equation}
  \label{eq:mu}
  \mu_\pm(\br) = \frac{\delta\mathcal{H}}{\delta c_\pm(\br)} 
  = \frac{\partial h}{\partial c_\pm} (\br) \pm q \psi(\br)
\end{equation}
from the free energy functional of the electrolyte given by (see,
e.g., Refs.~\cite{Dominguez2008,Podgornik2018,Stone2020}) 
\begin{eqnarray}
  \label{eq:H}
  \mathcal{H}[c_+(\br),c_-(\br),\psi(\br)]
  & =   
  & \int\limits_\mathrm{fluid}
  d^3\br\;\left[ 
    h(c_+,c_-) - \frac{1}{2} \epsilon |\nabla \psi|^2
    \right.
    \nonumber
  \\
  &
  & \!\!\!\!\!\! \!\!\!\!\!\! \!\!\!\!\!\! \!\!\!\!\!\! \!\!\!\!\!\!
    \!\!\!\!\!\! \!\!\!\!\!\! \!\!\!\!\!\!
    \left. \phantom{\frac{1}{2}} + q (c_+-c_-) \psi - 
    \sigma_s \, \mathbb{S}(\br_p) \, \psi \, \delta(|\br|-R) \right] \; .
\end{eqnarray}
Here, $\psi(\br)$ denotes the electric potential, and $h(c_+,c_-)$ is a local
free energy density that depends implicitly on temperature. A simple and frequently 
employed choice is the ideal gas form,
\begin{equation}
\label{eq:h_id_gas}
\beta \,h (c_+,c_-) = c_+ \left(\ln \frac{c_+}{c_0} - 1\right)
+ c_- \left(\ln \frac{c_-}{c_0} - 1\right) \, .
\end{equation} 
In such case, \eqs{eq:cons_ion}, \eref{eq:f}, and \eref{eq:mu} render
the usually employed Nernst--Planck equations. Although the formalism
to be presented can be applied in full generality, for instance by
addressing a local free energy $h$ that accounts for steric effects, we will also use \eq{eq:h_id_gas} both for reasons of simplicity and for straightforward comparison with previous studies.

The equations are supplemented by the boundary conditions at infinity, 
\begin{subequations}
\label{eq:BCs_diff}
\begin{equation}
\label{eq:BC_inf_cs}
\mu_\pm(|\br| \to \infty) \to \mu_0\,,
\quad
c_\pm(|\br| \to \infty) \to c_0\,, 
\end{equation}
and at the surface of the particle, where the catalytic activity is modeled as a 
current along the direction normal to the particle:
\begin{equation}
\label{eq:BC_part_cs}
\be_r \cdot \bj_+(\br_p)
= \be_r \cdot \bj_-(\br_p)
= \mathcal{A} \,\mathbb{A}(\br_p)\,.
\end{equation}
\end{subequations}

\noindent\textit{Electric field.--} The potential $\psi(\br)$ in the
fluid is determined from the minimization of the free energy
functional with respect to $\psi(\br)$, i.e., by solving
$\delta\mathcal{H}/\delta \psi(\br) = 0$. This renders the Poisson equation,
\begin{equation}
\label{eq:Poisson}
\nabla^2 \psi(\br) = -\frac{q}{\epsilon}[c_+(\br)-c_-(\br)]\,,
\end{equation}
and the boundary condition at the surface of the particle,
\begin{subequations}
\label{eq:BCs_pot} 
\begin{equation}
\label{eq:BC_part_psi}
\be_r \cdot \nabla \psi(\br_p) = -\frac{\sigma_s}{\epsilon}\, \mathbb{S}(\br_p)\,,
\end{equation}
associated to a high dielectric constant. 
These are supplemented by the boundary condition at infinity,
\begin{equation}
\label{eq:BC_inf_psi}
\psi(|\br| \to \infty) \to 0\,.
\end{equation}
\end{subequations}

\noindent\textit{Hydrodynamics and the motion of the particle.--} The flow
is determined by the incompressible Stokes equations, i.e., by the
mechanical balance between the fluid stresses and the body force
density, $\bff(\br) := \bff_+(\br) + \bff_-(\br)$, acting on the
solution, complemented by boundary conditions at infinity (quiescent
fluid) and at the surface of the particle (no slip) \cite{SM}. The
translational and rotational velocities, $\bV$ and $\bOmega$
respectively, of the overdamped motion of the particle can be obtained
from the condition of mechanical isolation of the composed system
``particle+solution'' by using the Lorentz reciprocal theorem, without
the need to compute first the velocity field in the fluid
\cite{Teubner1982,Seifert2012,Koplik2013,DPRD20,DP24}:
\begin{subequations}
  \label{eq:VOm}
\begin{eqnarray}
  \label{eq:V}
  \bV
  & =
  & \frac{1}{6\pi \eta R} \int\limits_\mathrm{fluid}
  d^3\br \; 
    \mathsf{K}(\br) \cdot \bff(\br),
  \\
  \label{eq:Omega}
  \bOmega
  & =
  & \frac{1}{8 \pi \eta R^3} \int\limits_\mathrm{fluid}
    d^3\br \;\mathbf{K}(\br) \times \bff(\br),
\end{eqnarray}
\end{subequations}
where the (in general non-symmetric) tensorial kernel
$\mathsf{K}(\br)$ and the vectorial kernel $\mathbf{K}(\br)$ are
determined solely by the geometrical shape of the particle
\cite{BrennerBook,KiKa91}. These fields incorporate the
incompressibility constraint as ($\mathsf{K}^\dagger$ is the transposed tensor)
\begin{equation}
  \label{eq:incomp}
  \nabla\cdot\mathsf{K}^\dagger=0 ,
  \qquad
  \nabla\times\bK=0 .
\end{equation}
This ensures that only the solenoidal component of $\bff$ contributes
in \eqs{eq:VOm} to the motion of the particle \cite{DPRD20,DP24}. For
the specific case of a spherical particle, they take the form
\begin{subequations}
  \label{eq:spherKernels}
\begin{equation}
  \label{eq:GV}
  \mathsf{K} (\br) =
  \left[ 
    \frac{1}{4} \left(\frac{R}{r}\right)^3
    + \frac{3R}{4r} - 1
  \right] \mathsf{I}
  + \frac{3R}{4r} \left[ 1 - \left(\frac{R}{r}\right)^2 \right]
  \be_r \be_r\,,
\end{equation}
\begin{equation}
  \label{eq:GOmega}
  \mathbf{K} (\br) = \left[ \left(\frac{R}{r}\right)^3 - 1 \right] r\,\be_r ,
\end{equation}
\end{subequations}
in spherical coordinates with origin at the center of the sphere,
where $\be_r$ is the unit radial vector and $\mathsf{I}$ is the
identity tensor. 

\noindent\textit{Quasi-homogenous regime and mapping to CICP.--} The
coupled \eqs{eq:cons_ion} and \eref{eq:Poisson} cannot be solved
analytically in general and progress can be made only after
approximations. We will focus on the case that the activity of the
particle and the surface charge are both very small, so that the
system is only weakly out of a spatially homogeneous equilibrium
state. Thus, the equations governing the ionic distribution will be
solved perturbatively to leading order in the deviations
$\delta c_\pm(\br):= c_\pm(\br) - c_0$ \footnote{This departs from the
somewhat more usual approach of perturbing around the equilibrium
state induced by a charged particle, which will be inhomogeneous. This
is again a simplification which reduces the algebraic burden without
sacrificing the physical picture.}.
In this approximation, Eqs.~(\ref{eq:mu}--\ref{eq:h_id_gas}) lead to
\begin{equation}
  \label{eq:deltamu}
  \delta\mu_\pm := \mu_\pm - \mu_0
  = \frac{1}{\beta c_0} {\delta c_\pm} \pm q\psi \, ,
\end{equation}
and one finds that the force density is given by \cite{SM}
\begin{equation}
\label{eq:def_fbar}
\bff(\br) 
= - \frac{1}{2} Q 
\nabla M 
- \frac{1}{4 \beta c_0} \nabla N^2 
\equiv - \frac{1}{2} Q(\br) \nabla M(\br)\,, 
\end{equation}
in terms of the local charge density (up to a factor $q$),
\begin{equation}
\label{eq:def_Q}
Q(\br) := \delta c_+ - \delta c_- = c_+-c_- \, ,
\end{equation}
the local contrast in chemical potential,
\begin{equation}
\label{eq:def_M}
M(\br) := \delta \mu_+ - \delta \mu_- = \mu_+ - \mu_-\, ,
\end{equation}
and the total concentration of ions
$N(\br) := \delta c_+ + \delta c_-$. This latter field has no effect
whatsoever on the motility due to \eqs{eq:incomp} for the
incompressibility constraint, in spite of it varying in space, as it
appears in \eq{eq:def_fbar} as an additive gradient that can be
dropped. Therefore, only charge imbalances matter in the linear
approximation. Upon defining the parameter
\begin{equation}
  \label{eq:meanD}
  \frac{1}{\meanD} := \frac{1}{D_+} - \frac{1}{D_-} ,
\end{equation}
and the Debye length
\begin{equation}
\label{eq:Debye_len}
\lambda_D := \sqrt{\frac{\epsilon}{2 q^2 \beta c_0}}\,,
\end{equation}
one finds \cite{SM} that \eqs{eq:cons_ion} lead to the boundary
value problem 
\begin{subequations}
\label{eq:bvp_M}
\begin{eqnarray}
\label{eq:Lap_M}
  \nabla^2 M(\br)
  & =
  & 0\,,
  \\
  \label{eq:BC_part_M}
  \be_r \cdot \nabla M(\br_p)
  & =
  & - \frac{\mathcal{A}}{\beta c_0 \meanD} \mathbb{A}(\br_p) \,,
  \\
  \label{eq:BC_inf_M}
  M(|\br| \to \infty)
  & =
  & 0\,,
\end{eqnarray}
\end{subequations}
while the electrostatic equation~(\ref{eq:Poisson}) yields
the boundary--value problem
\begin{subequations}
\label{eq:bvp_Q}
\begin{eqnarray}
\label{eq:Helm_Q}
  \nabla^2 Q(\br) 
  & =
  & \frac{1}{\lambda_D^{2}} Q(\br) \,,
  \\
\label{eq:BC_part_Q}
  \be_r \cdot \nabla Q(\br_p)
  & =
  & -\frac{\mathcal{A}}{\meanD}
    \mathbb{A}(\br_p) 
    + \frac{\sigma_s}{q\lambda_D^2}\mathbb{S}(\br_p) \,, 
  \\
\label{eq:BC_inf_Q}
  Q(|\br| \to \infty)
  & =
  & 0\,.
\end{eqnarray}
\end{subequations}
The equations~(\ref{eq:def_fbar}, \ref{eq:bvp_M}, \ref{eq:bvp_Q}) have
the same mathematical structure as the model for self-phoretic
motility by CICP \cite{DPRD20}, with 
$\lambda_D$ playing the role of the correlation length \cite{SM}.
This mathematical mapping, which is the main result of this work,
allows one to directly import the results for $\bV, \bOmega$ derived
by \rcite{DPRD20} to the current problem of ionic self-phoresis. It
provides a conceptually clear and physically insightful unifying
perspective on a number of previously reported results. A first,
immediate result is the absence of phoretic motility in this
approximation when the two ionic species have the same diffusivity
($\meanD^{-1}=0 \Rightarrow M=0 \Rightarrow \bff=0$)
\cite{Corato2020,Cruz2024}.

The solution of \eqs{eq:bvp_Q} can be written as the superposition
$Q(\br) = Q^{(\mathbb{A})} (\br) + Q^{(\mathbb{S})} (\br)$ \cite{SM},
where $Q^{(\mathbb{S})} (\br)$ is the charge distribution induced
solely by the particle surface charge (i.e., the charge distribution
in equilibrium), while $Q^{(\mathbb{A})} (\br)$ is the charge
distribution due to activity alone --- since $\mathbb{A}$ enters
\eqs{eq:bvp_Q} on equal footing with $\mathbb{S}$, the activity
appears to renormalize the particle's charge.  Accordingly, the
phoretic velocities are, via Eqs.~(\ref{eq:VOm}, \ref{eq:def_fbar}),
similarly decomposed. By expanding the activity in spherical
harmonics,
\begin{equation}
  \label{eq:alm}
  \mathbb{A}(\br_p) = \sum_{\ell=0}^\infty \sum_{m=-\ell}^{\ell}
  a_{\ell m} Y_{\ell m}(\theta,\varphi) ,
\end{equation}
and likewise for the other surface field $\mathbb{S}(\br_p)$ (with
coefficients $s_{\ell m}$), one can write \cite{DPRD20,SM}
  \begin{subequations}
    \label{eq:phoreticVel}
    \begin{eqnarray}
      \label{eq:tildeV}
      \bV
      & = 
      & \sum_{\ell m, \ell' m'}
        \left[ \left(\frac{R}{\lambda_D}\right)^2
        V^{(\mathbb{S})} \, s_{\ell m} 
        - V^{(\mathbb{A})} \, a_{\ell m} 
        \right] a_{\ell' m'} \,
        \\
      & & \times \left[
      g^{\perp}_{\ell \ell'}\left(\frac{\lambda_D}{R}\right)
      \bG^{\perp}_{\ell m; \ell' m'}
      + g^{\parallel}_{\ell \ell'} \left(\frac{\lambda_D}{R}\right)
      \bG^{\parallel}_{\ell m; \ell' m'}
          \right] ,
          \nonumber 
      \\
    \label{eq:tildeomega}
    \bOmega
      & =
      & 
      \sum_{\ell m, \ell' m'}
        \left(\frac{R}{\lambda_D}\right)^2
        \Omega^{(\mathbb{S})} \,
        s_{\ell m} a_{\ell' m'} \, 
      g^{\tau}_{\ell \ell'} \left(\frac{\lambda_D}{R}\right) \,
      \bG^{\tau}_{\ell m; \ell' m'}\, ,
    \end{eqnarray}
    with the characteristic velocity scales 
    \begin{equation}
      \label{eq:defVscale}
      V^{(\mathbb{A})} :=
      \frac{q^2 R^5 \mathcal{A}^2}{6 \pi \eta \epsilon \meanD^2} ,
      \quad
      V^{(\mathbb{S})} :=
      \frac{q R^3 \mathcal{A} \sigma_s}{6 \pi \eta \epsilon \meanD} ,
      \quad
      \Omega^{(\mathbb{S})} := \frac{3
        V^{(\mathbb{S})}}{4 R} .
    \end{equation}
  \end{subequations}
  In these expressions, the dimensionless functions $g^{\parallel}$,
  $g^{\perp}$, $g^{\tau}$ encode the dependence on the bulk
  concentration $c_0$ through the Debye length, and the dimensionless
  vectors $\bG^{\perp}$, $\bG^{\parallel}$, $\bG^{\tau}$, which are
  purely geometrical factors independent of any system parameters,
  vanish for certain combinations of modes $\{\ell,m\}$ and
  $\{\ell',m'\}$ --- accordingly, they imply ``selection rules'',
  i.e., those modes do not contribute to the phoretic velocities:
\begin{subequations}
  \label{eq:selection}
  \begin{eqnarray}
    \!\!\!\!\left.
    \begin{array}[c]{c}
      \bG^{\perp}_{\ell m; \ell' m'} = 0
      \\
      \bG^{\parallel}_{\ell m; \ell' m'} = 0
    \end{array}
    \right\}
    & \textrm{ if }
    & \left\{
      \begin{array}[c]{l}
        \ell-\ell' \neq \pm 1 , \mathrm{or}
        \\
        m+m' \neq 0, \pm 1 ,        
      \end{array}
      \right.
    \\
    \!\!\!\!
    \bG^{\tau}_{\ell m; \ell' m'} = 0
    &  \textrm{ if }
    & \left\{
      \begin{array}[c]{l}
        \ell-\ell' \neq 0 , \mathrm{or}
        \\
        \ell=\ell'=0 , \mathrm{or}
        \\
        m = m', \mathrm{or}
        \\
        m+m' \neq 0, \pm 1 .
      \end{array}
    \right.
  \end{eqnarray}  
\end{subequations}
The index $\parallel$ denotes the contributions coming from the components
of the force~(\ref{eq:def_fbar}) tangential to the surface of the
spherical particle, while $\perp$ pertains to the normal
component. The angular velocity $\bOmega$ only receives contributions
from the tangential components \cite{DPRD20}.

\noindent\textit{Discussion.--} For the physical interpretation of
\eqs{eq:phoreticVel}, it is useful to separately highlight
\cite{DoPo22} the \textit{source of phoresis}, associated to the
problem~(\ref{eq:bvp_M}) describing how the system is driven out of
equilibrium (in this case, the only source is activity), and the
\textit{mechanisms of phoresis}, associated to the
problem~(\ref{eq:bvp_Q}), which concerns how particle motion emerges. 

Pertaining to translation, there are two contributions in $\bV$,
similar in mathematical structure but with different physical meaning
and dependence on the system parameters. Thus, the
``surface-charge--driven mechanism'' corresponding to
$V^{(\mathbb{S})}$ follows from the coupling between the equilibrium
charge distribution induced by the particle charge (the piece
$Q^{(\mathbb{S})}(\br)$ of the solution to \eqs{eq:bvp_Q}), and the
departure from equilibrium driven by activity. This component of the
motility mechanism can be captured by linear-response theory and may
eventually be interpreted as ``phoresis in self-generated
gradients''. On the contrary, the contribution associated to
$Q^{(\mathbb{A})}(\br)$ and $V^{(\mathbb{A})}$, which depends entirely
on activity and was first identified in
Refs.~\cite{Corato2020,DPRD20}, precludes a similar interpretation due
to the specific bi-linear dependence on the activity rate
$\mathcal{A}$; that is, it cannot be derived in the classic framework
of \rcite{Prieve1984_JFM}. This contribution is precisely the
``correlation--driven mechanism'' of chemophoresis
\cite{DPRD20,DoPo22}, so termed because it vanishes in the absence of
correlations (i.e., when $\lambda_D\to 0$, see, c.f.,
\eqs{eq:g_small_xi}).

As for the rotational velocity $\bOmega$, it highlights the conceptual
difference between the two mechanisms, which manifests qualitatively
by the lack of a contribution $\Omega^{(\mathbb{A})}$. As shown by
\rcite{DPRD20}, this is due to the fact that chirality is not broken
when both the source and the mechanism of phoresis (as defined above)
have the same origin --- the chemical activity, in this case. On the
other hand, in the generic case that the origins are unrelated, as in
the case of activity vs.~surface charge, a preferred direction of
rotation emerges for the particle to attempt to restore equilibrium,
which explains the contribution $\Omega^{(\mathbb{S})}$ that is linear
in the activity and in the charge. Accordingly, in experimental studies
the rotational phoresis could play a key role in disentangling the
contributions of the two mechanisms (surface-charge--driven
vs.~correlation--driven), as in this quasi--homogeneous approximation
$\bOmega$ is the telltale signature of, and it is in whole attributed
to, the former.

The selection rules~(\ref{eq:selection}) provide constraints on how
motility emerges. Focusing on the translational motion, it occurs
through the coupling of successive multipoles of source and mechanism,
respectively (activity and surface charge for $V^{(\mathbb{S})}$, or
just activity for $V^{(\mathbb{A})}$).  This ensures that at least one
of the two is polar, i.e., there must exist a ``fore-aft'' asymmetry
in the particle surface properties. The simplest choices are, e.g., a
monopole plus dipole of activity ($a_{\ell m}=0$ if $\ell \geq 2$),
with either a monopolar surface charge ($s_{\ell m}=0$ if
$\ell \neq 0$) as was done in \rcite{Corato2020}, or with a monopole
plus dipole of surface charge ($s_{\ell m}=0$ if $\ell \geq 2$) as was
the case in \rcite{Cruz2024}. But, obviously, \eqs{eq:phoreticVel}
allow one to predict very many other choices of activity patterns for
which motion would occur, e.g., activity patterns missing the
monopolar and dipolar components but possessing a quadrupole component
which would couple with the dipole of the surface charge. This
observation also clarifies that \rcite{Brown2017} missed the
``correlation--driven'' contribution (quadratic in the activity) not
because the particle considered was not a net source of ion pairs
($a_{00}=0$), but rather because activity was modeled \emph{with a
  single} multipole.

The phoretic velocities depend on the background ion concentration
$c_0$ through the ratio $\lambda_D/R$ as encoded by the $g$
functions. The dependence may be quite complex, but robust features
emerge in the limiting cases. When $\lambda_D/R \to 0$ (the so-called
``thin--(Debye)--layer approximation'' \cite{Brown2017,Koplik2013,
  Prieve1984_JFM,Dukhin1993,Michelin2014,Derjaguin,Anderson1989,DPRD20},
because the force field $\bff(\br)$ is non-vanishing only in a thin
layer lying on the particle surface), one gets the universal behavior
\cite{SM}
\begin{equation}
  \label{eq:g_small_xi}
  g^{\parallel}_{\ell \ell'} \approx \frac{3}{2}
  \left(\frac{\lambda_D}{R}\right)^5 ,
  \quad
  g^{\perp}_{\ell \ell'} \approx 2 \frac{\lambda_D}{R} g^{\parallel}_{\ell \ell'},
  \quad
  g^{\tau}_{\ell \ell'} \approx 2 g^{\parallel}_{\ell \ell'} .
\end{equation}
That is, in this limit the phoretic velocities are dominated by the
tangential components of the force and they are predicted to vanish
with increasing electrolyte concentration as
$V,\Omega \sim c_0^{-3/2}$ if $V^{(\mathbb{S})} \neq 0$, or as
$V \sim c_0^{-5/2}$ if $V^{(\mathbb{S})} = 0$. A vanishing velocity
agrees \footnote{Under the proviso that, irrespective of the nature of
  the ions, the addition of salts can be interpreted to affect only
  the value of $c_0$, and thus of $\lambda_D$.}  with experimental
observations~\cite{Ebbens2014,Palacci2013,BrPo14}. In the opposite
limit, $\lambda_D/R \to \infty$, the phoretic velocities
$\bV, \bOmega$ reach finite values that are independent of $\lambda_D$
(thus of $c_0$) \cite{SM}. It is noteworthy that both limiting
behaviors hold regardless of the specific combination of activity and
surface charge multipoles, and thus extend the results reported for
specific choices \cite{Brown2017,Bayati2016,Corato2020,Cruz2024}.

\begin{figure}[!tb]
  (a) $V^{(\mathbb{S})}/V^{(\mathbb{A})} = 10^{-2}$
  \\
  \includegraphics[width=0.95\columnwidth]{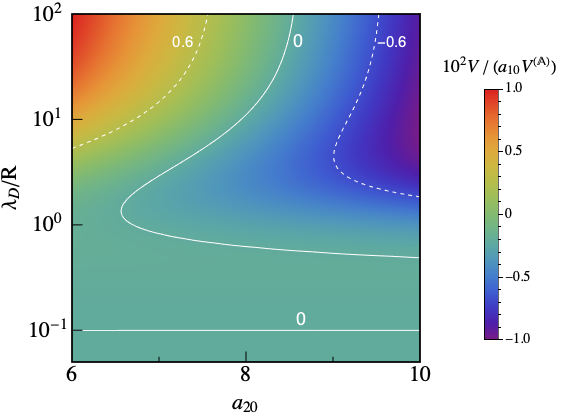}
  \\
  $\;$
  \\
  (b) $a_{20}=7$
  \\
  \includegraphics[width=0.9\columnwidth]{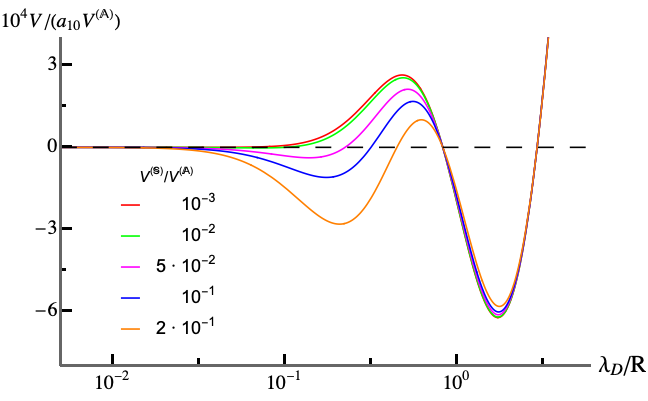}
  \caption{\label{fig:example} Translation velocity $V=\be_z\cdot\bV$
    as a function of the scaled Debye length $\lambda_D/R$, the
    quadrupolar moment $a_{20} (\equiv s_{20})$, and the ratio
    $V^{(\mathbb{S})}/V^{(\mathbb{A})}$ (see the main text for
    details). It is shown in (a) at fixed
    $V^{(\mathbb{S})}/V^{(\mathbb{A})}=10^{–2}$ (white curves are
    contour lines of constant $V$) and in (b) at fixed
    $a_{20}=s_{20}=7$.  }
\end{figure}
Beyond these limiting cases, the different dependence on the system
parameters of the two contributions to $\bV$, but also the combination
of multipole moments in each of them, can be expected to lead
generically to changes in the direction of motion upon varying $c_0$
and $R$ \cite{Brown2017,Corato2020,Cruz2024}, even multiple times
\cite{DPRD20}. Figure~\ref{fig:example} provides an illustrative
example of the complexity emerging when going beyond the usually
considered configuration of monopole plus dipole; e.g., we consider a
model system with $a_{00}=s_{00} = 1$, $a_{10}=s_{10} \neq 0$,
$a_{20}=s_{20} \neq 0$, all other multipoles being zero \footnote{This
  would correspond to, e.g., the lowest orders expansions for a
  typical Janus particle, for which the activity is due to an
  axisymmetric decoration by a catalyst of the surface of a
  homogeneous spherical core, so that activity and surface charge
  distributions are expected to be correlated.}, so that
$\bV=\be_z V$, $\bOmega=0$ according to \eqs{eq:phoreticVel}, and
choose realistic values of the ratio
$V^{(\mathbb{S})}/V^{(\mathbb{A})} = \sigma_s \Delta /q R^2
\mathcal{A}$ \footnote{Based on the values for $\mathcal{A}$ and
  $\sigma_s$ provided in \rcite{Cruz2024} and $q = e$, we estimated
  $|V^{(\mathbb{S})}/V^{(\mathbb{A})}|\approx 10^{-2}$ for a particle
  of radius $R = 500~\mathrm{nm}$.}.
Panel (a) shows that, upon varying the Debye length, the translational
velocity can change sign once (e.g., at $a_{20} = 6$), twice (at
$a_{20} \approx 6.5$), or three times (e.g., at $a_{20} = 7$) due to
the competition between the two mechanisms in \eq{eq:tildeV} (surface-charge--driven
  vs.~correlation--driven). In panel (b) we illustrate how this behavior depends on the
  ratio $V^{(\mathbb{S})}/V^{(\mathbb{A})}$ for a fixed $a_{20} = 7$. Moreover, it highlights that the values of the Debye length where this competition is more significant lies in the experimentally accessible range (from
 $R = 1~\mathrm{\mu m}$ and $\lambda_D/R = 10^{-2}$, corresponding to
  $\sim 1\;\mathrm{mM}$ salt concentration, to $R = 100~\mathrm{nm}$
  and $\lambda_D/R = 3$ for deionized water). This behavior bears similarity with certain experimental observations,
both in self-phoresis \footnote{Under the same proviso as in
  \cite{Note3}, this motility mechanism may consistently
  capture, as noted in \rcite{Brown2017}, the qualitative features of
  the velocity dependence on $\lambda_D/R$ observed in experiments
  with Pt-covered silica or polystyrene Janus spheres upon varying
  their size or adding small amounts of salts
  \cite{BrPo14,ebbens2012PRE,Ebbens2014}. This provides an alternative
  to the models based on an electrokinetic current through the Pt film
  \cite{Ebbens2014,Liverpool2017,Tasinkevych2019} or a very complex
  surface reaction \cite{Ebbens2014, ebbens2012PRE}.}\nocite{ebbens2012PRE}
and in classical phoresis in electrolyte gradients \cite{Prieve1988}. 

\noindent\textit{Conclusion.--} We have shown that an exact mathematical
mapping exists between a model of ionic self-phoresis 
and the model of correlation--induced self-phoresis introduced in
\rcite{DPRD20}. This conceptual connection builds a consistent
unifying perspective for previously derived results, provides
several additional theoretical predictions, and reveals insightful 
similarities between ionic self-phoresis and self-chemophoresis.

A number of directions for further exploration emerge. A study beyond
the quasi-homogeneous approximation, e.g., layering effects due to a
finite ion size, should reveal to what extent the mapping holds beyond
this simplification. An extension to the case of multiple types of ion
pairs in the electrolyte, including electrolytes with asymmetric ions,
would be also useful in order to address realistic experimental
configurations. Finally, it would be interesting to examine other
mechanisms of ionic activity that involve either the release of a
single species of ions, e.g., the redox reactions in bi-metallic
structures \cite{Moran2011}, or an ion exchange with the electrolyte,
as in the case of the Nafion \cite{Esplandiu2020,Esplandiu2022} or
resin \cite{Palberg2017,Palberg2018} particles.

\begin{acknowledgments}
\label{Acknowledgments}
The authors 
acknowledge financial support through grants ProyExcel\_00505 funded
by Junta de Andaluc{\'i}a, and PID2021-126348NB-I00 funded by
MCIN/AEI/10.13039/501100011033 and ``ERDF A way of making
Europe''. M.N.P.~also acknowledges support from Ministerio de
Universidades through a Mar{\'i}a Zambrano grant.
\end{acknowledgments}

%
%

\end{document}


\title{Ionic self-phoresis maps onto correlation--induced
  self-phoresis
  \\
  Supplemental Material\footnote{Here, equations of the main text are
    referenced by their respective numbers. Equations appearing only
    in the Supplemental Material are cited by a combination of the
    section number (in Roman numerals) and the equation number within
    the section (in Arabic numerals).}}

\author{Alvaro Dom\'\i nguez}
\email{\texttt{dominguez@us.es}}
\affiliation{
  F{\'i}sica Te{\'o}rica, Universidad de Sevilla, Apdo.~1065, 
41080 Sevilla, Spain}
\affiliation{Instituto Carlos I de F{\'i}sica Te{\'o}rica y Computacional,
  18071 Granada, Spain}

\author{Mihail N.~Popescu}
\email{\texttt{mpopescu@us.es}}
\affiliation{
  F{\'i}sica Te{\'o}rica, Universidad de Sevilla, Apdo.~1065, 
41080 Sevilla, Spain}

\date{\today}

\maketitle



\section{Hydrodynamics}

The flow $\bu(\br)$ is determined by the mechanical balance
between the body force density acting on the solution, see \eq{eq:f},
\begin{equation}
  \bff(\br) = \bff_+(\br) + \bff_-(\br)
  =   \mbox{} - c_+ (\br) \nabla\mu_+(\br)
  - c_- (\br) \nabla\mu_- (\br),
\end{equation}
and the fluid stresses. This is expressed by the Stokes equation for
incompressible flow,
\begin{equation}
  \label{eq:stokes}
  \eta\nabla^2\bu - \nabla p + \mathbf{f} = 0 ,
  \qquad
  \nabla\cdot\bu=0 ,
\end{equation}
where $\eta$ is the viscosity of the solution and $p$ is the pressure
field enforcing incompressibility. The hydrodynamic flow is subject to
the boundary conditions of quiescent solution at infinity,
\begin{subequations}
\label{eq:BCs_flow}
\begin{equation}
\label{eq:BC_inf_flow}
\bu(|\br| \to \infty) \to 0\, ,
\end{equation}
which sets the rest frame with respect to which the velocities are
measured, and of no slip at the surface of the particle, which
translates with velocity $\bV$ and rotates with angular velocity
$\bOmega$,
\begin{equation}
\label{eq:BC_part_flow}
\bu(\br_p) = \bV + \bOmega \times \br_p\,.
\end{equation}
\end{subequations}
Finally, the yet unknown translational and rotational velocities $\bV$
and $\bOmega$ of the overdamped motion of the particle are fixed by
the requirement that the system ``particle + fluid'' does not
experience external forces or torques.

\section{\label{sec:linear} Details of the calculations in the quasi-homogeneous regime}

The procedure is very similar to the one carried out in
Ref. \cite{DPRD20}.  We start with the body force
density~(\ref{eq:f}), which, up to second--order in the small
deviations from homogeneity, is given by
\begin{subequations}
\begin{equation}
\label{eq:fpm_lin}
  \bff_\pm
  = -(c_0 + \delta c_\pm) \nabla(\mu_0 + \delta \mu_\pm)
  = -c_0 \nabla \delta \mu_\pm - \delta c_\pm \nabla \delta \mu_\pm \, 
\end{equation}
\begin{equation}
  \Rightarrow\quad \bff = -c_0 \nabla \left( \delta \mu_+ + \delta \mu_- \right)
  - \delta c_+ \nabla \delta \mu_+
  - \delta c_- \nabla \delta \mu_- \, .
\end{equation}
\end{subequations}
It is immediately apparent that, in the flow problem, it is necessary
to go to the second order because at the first order, $\bff$ is a
gradient, and thus cannot contribute to the fluid motion.
But in the diffusion equations~(\ref{eq:cons_ion}), the expansion of
$\bff_\pm$ to first order is sufficient, and it leads to Laplace
equations for the chemical potential deviations, 
\begin{subequations}
\label{eq:lin_bvp_mupm}
\begin{equation}
\label{eq:Lap_mupm}
\nabla^2 \delta \mu_\pm = 0\,,
\end{equation}
with boundary conditions at the surface of the particle following from
\eqs{eq:BC_part_cs},
\begin{equation}
  \label{eq:BC_part_lin_mupm}
  \be_r \cdot \nabla \delta \mu_\pm(\br_p) = 
  -\frac{\cal A}{c_0 \Gamma_\pm} \mathbb{A}(\br_p)\,,
\end{equation}
and at infinity, corresponding to the particle reservoir, see
\eq{eq:BC_inf_cs},
\begin{equation}
\label{eq:BC_inf_lin_mupm}
\delta \mu_\pm(|\br| \to \infty) = 0\,.
\end{equation}
\end{subequations}

Similarly, for the electrostatic problem the first order expansion is
sufficient, and Eqs.~(\ref{eq:Poisson}, \ref{eq:BCs_pot}) render the
boundary--value problem
\begin{subequations}
\label{eq:lin_bvp_pot} 
\begin{equation}
\label{eq:Poisson_lin_psi}
\nabla^2 \psi = -\frac{q}{\epsilon} \left( \delta c_+ - \delta c_-
\right) \,,
\end{equation}
\begin{equation}
\label{eq:BC_part_lin_psi}
\be_r \cdot \nabla \psi(\br_p) = -\frac{\sigma_s}{\epsilon}\,
\mathbb{S}(\br_p)\, ,
\end{equation}
\begin{equation}
\label{eq:BC_inf_lin_psi}
\psi(|\br| \to \infty) \to 0\,.
\end{equation}
\end{subequations}

The ``excess'' concentrations $\delta c_\pm$ follow from expanding at
linear order the chemical potential~(\ref{eq:mu}) with the ideal gas
approximation~(\eq{eq:h_id_gas}):
\begin{equation}
\label{eq:lin_exp_mu}
  \delta \mu_\pm
  = \delta c_\pm \frac{\partial^2 h}{\partial c_\pm^2}\left(c_0,
    c_0\right)
    + \delta c_\mp \frac{\partial^2 h}{\partial c_+ \partial
    c_-}\left(c_0, c_0\right) \pm q \psi
  = \frac{1}{\beta c_0} \delta c_\pm \pm q \psi\,,
\end{equation}
which is \eq{eq:deltamu}. One notices immediately the relation
\begin{equation}
\label{eq:dc_sum} 
N(\br) := \delta c_+ + \delta c_- = \beta c_0 (\delta \mu_+ + \delta \mu_-)\,,
\end{equation}
for the deviations from homogeneity of the total concentration of
ions.  With $\delta \mu_\pm(\br)$ known as the solution of the
boundary--value problem~(\ref{eq:lin_bvp_mupm}), the
relations~(\ref{eq:lin_exp_mu}) provide $\delta c_\pm(\br)$; by
plugging these into \eqs{eq:lin_bvp_pot} the boundary--value problem
for the electric potential $\psi(\br)$ can be solved, and all the
relevant functions are thus determined.

However, for the purpose of just calculating the phoretic velocities
of the particle, not all of the above steps are necessary. As noticed,
the first order terms in the expression~(\ref{eq:fpm_lin}) for the
force do not make contributions to the velocity. At the second order,
we have
\begin{eqnarray}
  \bff
  & =
  & \bff_+ + \bff_-
  =
  -(\delta c_+ \nabla \delta \mu_+ + \delta c_- \nabla \delta \mu_-) 
  =
  -\frac{1}{2}(\delta c_+ + \delta c_-)\nabla (\delta \mu_+ + \delta
    \mu_-)
  - \frac{1}{2}(\delta c_+ - \delta c_-)\nabla (\delta \mu_+ - \delta
  \mu_-)
  \nonumber
  \\
  \label{eq:force2}
  & =
  & - \frac{1}{4 \beta c_0} \nabla \left(N^2 \right)
  - \frac{1}{2} Q\nabla M ,
\end{eqnarray}
after using the definitions~(\ref{eq:dc_sum}, \ref{eq:def_Q},
\ref{eq:def_M}). The first term in the final expression is a perfect
gradient, and thus irrelevant for the motion of the particle;
accordingly, after dropping it we arrive at the (for our purposes
equivalent) expression~(\ref{eq:def_fbar}) for the body force density
responsible for the particle motility.

By subtracting the two problems contained in \eqs{eq:lin_bvp_mupm}
(one for $\delta \mu_-$, another one for $\delta \mu_+$), one
immediately obtains the boundary--value problem~(\ref{eq:bvp_M}) obeyed
by $M(\br)$. Likewise, by subtracting the two \eqs{eq:lin_exp_mu}, one
obtains the relation
\begin{equation}
  \label{eq:Q_M_psi}
  2 q \psi(\br) = M(\br) - \frac{1}{\beta c_0} Q(\br) .
\end{equation}
And then, the electrostatic boundary--value
problem~(\ref{eq:lin_bvp_pot}), complemented by \eqs{eq:bvp_M}, leads
to the problem~(\ref{eq:bvp_Q}) for the charge concentration
field. Due to the superposition of sources in \eq{eq:BC_part_Q}, one
can write the solution as
\begin{equation}
  \label{eq:QAS}
  Q(\br) = Q^{(\mathbb{A})} (\br) + Q^{(\mathbb{S})} (\br)\,, 
\end{equation}
where each field satisfies a different problem:
\begin{subequations}
\label{eq:bvp_QA}
\begin{eqnarray}
  \label{eq:Helm_QA}
  \nabla^2 Q^{(\mathbb{A})}(\br) 
  & =
  & \frac{1}{\lambda_D^{2}} Q^{(\mathbb{A})}(\br) \,,
  \\
  \label{eq:BC_part_QA}
  \be_r \cdot \nabla Q^{(\mathbb{A})}(\br_p)
  & =
  & -\frac{\mathcal{A}}{\meanD}
    \mathbb{A}(\br_p) \,, 
  \\
  \label{eq:BC_inf_QA}
  Q^{(\mathbb{A})}(|\br| \to \infty)
  & =
  & 0\,,
\end{eqnarray}
\end{subequations}
and
\begin{subequations}
\label{eq:bvp_QS}
\begin{eqnarray}
  \label{eq:Helm_QS}
  \nabla^2 Q^{(\mathbb{S})}(\br) 
  & =
  &
    \label{eq:BC_part_QS}
    \frac{1}{\lambda_D^{2}} Q^{(\mathbb{S})}(\br) \,,
  \\
  \be_r \cdot \nabla Q^{(\mathbb{S})}(\br_p)
  & =
  & \frac{\sigma_s}{q\lambda_D^2}\mathbb{S}(\br_p) \,, 
  \\
  \label{eq:BC_inf_QS}
  Q^{(\mathbb{S})}(|\br| \to \infty)
  & =
  & 0\,,
\end{eqnarray}
\end{subequations}
respectively.

\section{\label{sec:map} Mapping between ionic self-phoresis and CICP}

The relevant ingredients of the model introduced in \rcite{DPRD20} for
CICP are the concentration field $n(\br)$ of a chemical, its chemical
potential $\mu(\br)$ and the associated free energy,
\begin{equation}
  \mathcal{H}_\mathrm{CICP}[n] = \int\limits_\mathrm{fluid}
  d^3\br\;\left[ 
    h(n) + \frac{1}{2} \lambda^2 |\nabla n|^2
  \right] ,
\end{equation}
where $\lambda$ is a parameter related to the correlation length (see
\eq{eq:CICPcorr}). Notice that, unlike with the free energy employed
in \eq{eq:H}, the correlations are modeled through a square--gradient
term in the concentration without the mediation of a field like the
electric potential $\psi(\br)$. The quasi-homogeneous approximation is
obtained by linearizing around the equilibrium state $n(\br)=n_0$, in
which the correlation length is given as
\begin{equation}
  \label{eq:CICPcorr}
  \xi = \frac{\lambda}{\sqrt{h''(n_0)}} .
\end{equation}
The following table provides, in a side-by-side form, the mapping
between this model for CICP and the model presented in the main text
for ionic chemophoresis.


\begin{center}
\begin{tabular}{c c c c c c}
  \begin{tabular}[c]{c}
      eq.~numbers in
      \\
      this work
  \end{tabular}
  & \hspace{.25in}
    \textbf{
  ionic chemophoresis}
  & \hspace{.5in}
  & 
    \begin{tabular}[c]{c}
      \textbf{
      correlation--induced}
      \\
      \textbf{chemophoresis (CICP)}
    \end{tabular}
  & \hspace{.25in}
  & \begin{tabular}[c]{c}
      eq.~numbers in
      \\
      Suppl.~Mat.~\cite{DPRD20}
    \end{tabular}
  \\
  \hline\hline
  \\
  (\ref{eq:def_fbar})
  &
  $\bff = -\dfrac{1}{2} Q\, \nabla M$
  & 
  & $\bff = -n_0 \Psi \, \nabla \mu$
  &
  & (III.31)
  \\
  \\
  \hline
  \\
  (\ref{eq:Lap_M})
  & 
  $\nabla^2 M = 0$
  & 
  & $\nabla^2\mu = 0$
  &
  & (II.5)
  \\
  \\
  (\ref{eq:BC_part_M})
  &
  $\be_r \cdot \nabla M(R \be_r) = - 
  \dfrac{\mathcal{A}}{\beta c_0 \Delta}
  \mathbb{A}(\be_r)$  
  & 
  & $\be_r \cdot \nabla \mu(R \be_r) = - \underbrace{\frac{q}{n_0
    \Gamma}}_{b_1} \mathbb{A}(\be_r)$
  &
  & (II.6)
  \\
  \\
  (\ref{eq:BC_inf_M})
  &
  $M(\br \to \infty) = 0$
  & 
  & $\mu(\br \to \infty) =\mu_0 $
  &
  & (II.6)
  \\
  \\
  \hline
  \\
  (\ref{eq:Helm_QA}, \ref{eq:Helm_QS})
  & $\nabla^2 Q^{(\mathbb{A}, \mathbb{S})} = \lambda_D^{-2} Q^{(\mathbb{A}, \mathbb{S})}$
  & 
  & $\nabla^2 \Psi = \xi^{-2} \Psi$
  &
  & (II.10)
  \\
  \\
  \begin{tabular}[c]{c}
    (\ref{eq:BC_part_QA})
    \\
    $\phantom{\Big|}$
    \\
    (\ref{eq:BC_part_QS})
  \end{tabular}
  &
    $\left.
    \begin{array}[c]{c}
    \be_r \cdot \nabla Q^{(\mathbb{A})}(R \be_r)
    = - \dfrac{\mathcal{A}}{\Delta} \mathbb{A}(\be_r)
    \\
    \\
    \be_r \cdot \nabla Q^{(\mathbb{S})}(R \be_r)
    = \dfrac{\sigma_s}{q \lambda_D^2}
    \mathbb{S}(\be_r)
  \end{array}
  \right\}$
  & 
  &
    $\be_r \cdot \nabla \Psi (R \be_r) = 
    \underbrace{\dfrac{q}{n_0^2 \Gamma} \left(\dfrac{\xi}{\lambda}\right)^2}_{b_2}
    \mathbb{A}(\be_r)$
  &
  & (II.12)
  \\
  \\
  \\
  (\ref{eq:BC_inf_QA}, \ref{eq:BC_inf_QS})
  & $Q^{(\mathbb{A}, \mathbb{S})}(\br \to \infty) = 0$
  & 
  & $\Psi(\br \to \infty) = 0$
  &
  & (II.11)
  \\
  \\
\hline\\\\
\end{tabular}
\end{center}
This table leads to the following ``translation rules'':
\begin{displaymath}
  \begin{array}{|c||c||c||c||c||c||c|}
    \hline
    \phantom{\Bigg|}
    \begin{array}[c]{c}
      \textbf{ionic}
      \\
      \textbf{chemophoresis}
    \end{array}
    & M
    &
      \begin{array}[c]{c|c}
        Q^{(\mathbb{A})}
        \phantom{\Bigg|}
        &
          Q^{(\mathbb{S})}    
      \end{array}
    & \lambda_D
    & \mathcal{A}/\beta c_0 \Delta
    &
      \begin{array}[c]{c|c}
        \mbox{} - \mathcal{A}/\Delta
        \phantom{\Bigg|}
        & \sigma_s/q\lambda_D^2
      \end{array}
    &
      \begin{array}[c]{c|c}
        \mathbb{A} = \sum_{\ell m} a_{\ell m} \, Y_{\ell m}
        \phantom{\Bigg|}
        &
          \mathbb{S} = \sum_{\ell m} s_{\ell m} \, Y_{\ell m}
      \end{array}
    \\
    \hline
    \phantom{\Bigg|}
    \textbf{CICP}
    &
    \mu-\mu_0
    & 2 n_0 \Psi
    & \xi
    & b_1
    & 2 b_2 n_0
    & \mathbb{A} = \sum_{\ell m} a_{\ell m} \, Y_{\ell m}
    \\
    \hline
  \end{array}
\end{displaymath}
The phoretic velocities for CICP can be represented with the same kind
of sums over modes as in \eqs{eq:phoreticVel} (see Eqs.~(III.37,
III.44) in \rcite{DPRD20})
\begin{equation}
  \label{eq:CICPvelocities}
  \bV = V_0 \sum_{\ell m,\ell' m'} a_{\ell m} a_{\ell' m'} \cdots ,
  \qquad
  \bOmega = \Omega_0 \sum_{\ell m,\ell' m'} a_{\ell m} a_{\ell' m'} \cdots ,
\end{equation}
with the characteristic scales (see Eqs.~(III.38, II.2))
\begin{equation}
  V_0 = b_1 b_2 n_0 \frac{R^3}{6\pi \eta} \left(\frac{R}{\xi}\right)^2 ,
  \qquad
  \Omega_0 = \frac{3 V_0}{4 R} .
\end{equation}
Therefore, making use of the translation table one arrives at the
expansions~(\ref{eq:tildeV}, \ref{eq:tildeomega}) with the scales
shown in \eq{eq:defVscale}:
\begin{subequations}
\begin{equation}
  V^{(\mathbb{A})} \;\longleftrightarrow\;  \mbox{} - V_0 = \mbox{} - \frac{\mathcal{A}}{\beta c_0 \Delta} \left(
    -\frac{\mathcal{A}}{2\Delta} \right)
  \frac{R^3}{6\pi \eta} \left(\frac{R}{\lambda_D}\right)^2
  = \frac{q^2 R^5 \mathcal{A}^2}{6\pi\eta\epsilon \Delta^2},
\end{equation}
\begin{equation}
  V^{(\mathbb{S})} \;\longleftrightarrow\; 
  \left(\frac{R}{\lambda_D}\right)^{-2} V_0
  = \left(\frac{R}{\lambda_D}\right)^{-2}
  \frac{\mathcal{A}}{\beta c_0 \Delta}
  \frac{\sigma_s}{2q \lambda_D^2}
  \frac{R^3}{6\pi \eta} \left(\frac{R}{\lambda_D}\right)^2
  = \frac{q R^3 \mathcal{A} \sigma_s}{6\pi\eta\epsilon \Delta} .
\end{equation}
\end{subequations}

\section{The phoretic velocities}

The expressions~(\ref{eq:phoreticVel}) follow from solving the
boundary--value problems~(\ref{eq:bvp_M}, \ref{eq:bvp_Q}) as
expansions in spherical harmonics (see \eq{eq:alm}), and evaluating
the integrals appearing in \eqs{eq:VOm} with \eqs{eq:spherKernels} for
the hydrodynamic kernels and \eq{eq:def_fbar} for the force
density. The details can be found in \rcite{DPRD20} and lead to
explicit expressions for the $g$ functions and the $\bG$
vectors. Specifically, one has (we here use the short-hand
  notations $\lambda \equiv \lambda_D/R$, $s \equiv r/R$)
\begin{subequations}
  \label{eq:gfunc}
\begin{equation}
  \label{eq:gr}
  g^{\perp}_{\ell \ell'} (\lambda) 
  := \int_1^\infty ds\; 
  \left( - \frac{1}{2s^3} + \frac{3}{2s} - 1
  \right) s^{-\ell'} \, c_\ell(s,\lambda)\,, 
\end{equation}
\begin{equation}
  \label{eq:gp}
  g^{\parallel}_{\ell \ell'} (\lambda) 
  := \int_1^\infty ds\; 
  \left( \frac{1}{4 s^3} + \frac{3}{4 s} - 1  \right)
  s^{-\ell'} \, c_\ell(s,\lambda)\,, 
\end{equation}
and
\begin{equation}
  \label{eq:gom}
  g^{\tau}_{\ell \ell'} (\lambda) :=
  \int_1^\infty ds\; 
  \left( \frac{1}{s^3} - 1 \right) s^{1-\ell'} \,
  c_\ell(s,\lambda) .
\end{equation}
\end{subequations}
One can recognize the pieces stemming from the hydrodynamic
kernels~(\ref{eq:spherKernels}), while the functions
\begin{equation}
\label{eq:def_littlec_ell}
c_\ell\left(s,\lambda\right) := \gamma_{\ell}(\lambda)\frac{K_{\ell +1/2}(s/\lambda)}{\sqrt{s}}\,,
\end{equation}
are the coefficients in the expansion in spherical harmonics of the
solution to \eqs{eq:bvp_Q}; the functions $K_n(x)$ are the modified Bessel
functions of the second kind of order $n$ \cite{abram,GrRy94},
and 
\begin{equation}
\label{eq:def_gamma}
\gamma_{\ell}(\lambda)=-\frac{\lambda^2}{(\ell+1)K_{\ell +1/2}(\lambda^{-1})+\lambda^{-1}K_{\ell-1/2}(\lambda^{-1})}\,.
\end{equation}
And one also has the vectors
  \begin{subequations}
    \label{eq:Gvect}
\begin{eqnarray}
  \label{eq:Gperp}
  \bG^\perp_{\ell m; \ell' m'} 
  & = 
  & \sqrt{(2\ell+1)(2\ell'+1)} 
    \left(
    \begin{array}[c]{ccc}
      1 & \ell & \ell' \\
      0 & 0 & 0 
    \end{array}
  \right)
  \;
  \\
  & 
  &     
    \times \left[ \be_z \, \left(
    \begin{array}[c]{ccc}
      1 & \ell & \ell' \\
      0 & m & m' 
    \end{array}
   \right) 
   - \frac{\be_x-i\be_y}{\sqrt{2}} \,
    \left(
    \begin{array}[c]{ccc}
      1 & \ell & \ell' \\
      1 & m & m' 
    \end{array}
    \right) 
    + \frac{\be_x+i\be_y}{\sqrt{2}} \,
    \left(
    \begin{array}[c]{ccc}
      1 & \ell & \ell' \\
      -1 & m & m' 
    \end{array}
  \right) 
\right]\,, ~~~~~~~~~~~
 \nonumber 
\end{eqnarray}
\begin{eqnarray}
\label{eq:Gpar}
  \bG^\parallel_{\ell m; \ell' m'} 
  & = 
  & - \frac{\sqrt{(2\ell+1)(2\ell'+1)}}{\ell'+1}
    \left(
    \begin{array}[c]{ccc}
      1 & \ell & \ell' \\
      0 & 0 & 0 
    \end{array}
  \right)
  \\
  & 
  & \times \left\{ m' 
    \left[ 
    \frac{\be_x+i\be_y}{\sqrt{2}} 
    \left(
    \begin{array}[c]{ccc}
      1 & \ell & \ell' \\
      -1 & m & m' 
    \end{array}
  \right)
   + \frac{\be_x-i\be_y}{\sqrt{2}} 
    \left(
    \begin{array}[c]{ccc}
      1 & \ell & \ell' \\
      1 & m & m' 
    \end{array}
  \right)
  \right]
  \right.
 \nonumber
 \\
&
& + \frac{\sqrt{\ell'(\ell'+1)-m'(m'+1)}}{\sqrt{2}}
  \left[ 
  \frac{\be_x-i\be_y}{\sqrt{2}}
  \left(
    \begin{array}[c]{ccc}
      1 & \ell & \ell' \\
      0 & m & m'+1 
    \end{array}
  \right)
 - \be_z \, \left(
    \begin{array}[c]{ccc}
      1 & \ell & \ell' \\
      -1 & m & m'+1 
    \end{array}
  \right)
 \right]
 \nonumber
  \\
&
& \left. + \frac{\sqrt{\ell'(\ell'+1)-m'(m'-1)}}{\sqrt{2}}
  \left[ 
    -\frac{\be_x+i\be_y}{\sqrt{2}}
  \left(
    \begin{array}[c]{ccc}
      1 & \ell & \ell' \\
      0 & m & m'-1 
    \end{array}
  \right)
 - \be_z \, \left(
    \begin{array}[c]{ccc}
      1 & \ell & \ell' \\
      1 & m & m'-1 
    \end{array}
  \right)
 \right]
 \right\}\,,
 \nonumber
 \end{eqnarray}
\begin{equation}
  \label{eq:Gom2}
  \bG^{\tau}_{\ell m; \ell'm'} =
  -\frac{i (-1)^m}{\ell+1} \delta_{\ell,\ell'}
    \left[ \be_z \, m' \delta_{m+m',0} 
    + \sqrt{\ell(\ell+1)+m \,m'}
    \left( 
    \frac{\be_x-i\be_y}{2} \, \delta_{m+m',-1} 
    + \frac{\be_x+i\be_y}{2} \, \delta_{m+m',1} 
    \right)
    \right] \,,
\end{equation}
  \end{subequations}
expressed in a Cartesian basis $\{\be_x, \be_y, \be_z\}$ and in terms
of the Wigner 3j symbols.

\subsection{Limiting behaviors in $\lambda =\lambda_D/R $}

\begin{figure}[!ht]
\includegraphics[width=0.3\columnwidth]{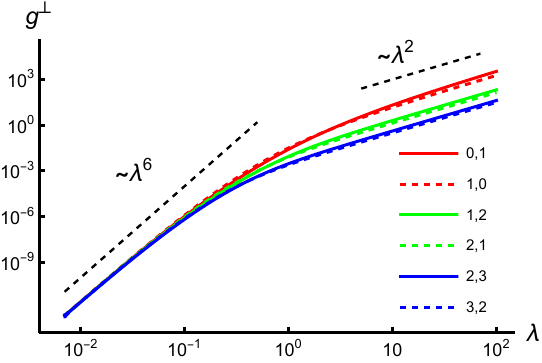}\hfill
\includegraphics[width=0.3\columnwidth]{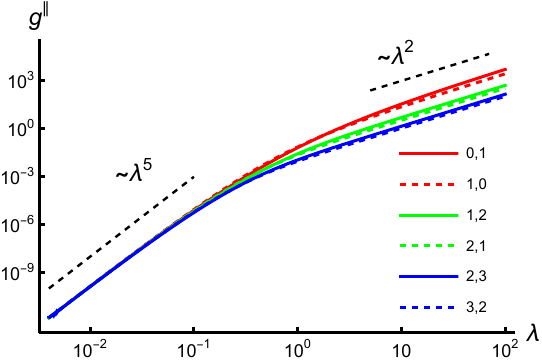}\hfill
\includegraphics[width=0.3\columnwidth]{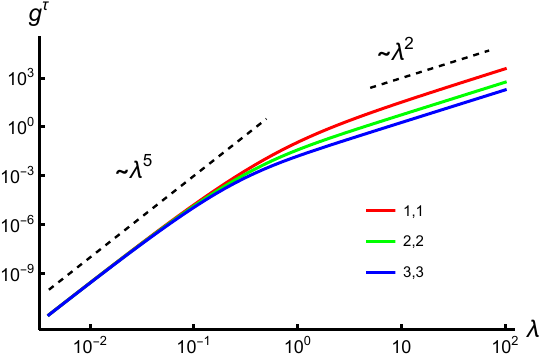}
\caption{\label{fig:g_functions} Log-log plots of the functions
  $g_{\ell \ell'}^{\perp}(\lambda), g_{\ell \ell'}^{\parallel}(\lambda),
  g_{\ell \ell'}^{\tau}(\lambda)$ for some choices of the pairs
  $(\ell, \ell')$ (as indicated by the labels) that fulfill the
  constraints~(\ref{eq:gselec}). Also shown are the asymptotic
  behaviors derived analytically.}
\end{figure}

The integrals appearing in the definitions of the $g$--functions can
be expressed analytically in terms of the exponential integral
function \cite{DPRD20}, and not all of them are needed: because of the
selection rules~(\ref{eq:selection}), the only functions that are
required are of the form
\begin{equation}
  \label{eq:gselec}
  g^\perp_{\ell,|\ell\pm 1|},
  \; g^\perp_{|\ell\pm 1|,\ell},
  \; g^\parallel_{\ell,|\ell\pm 1|},
  \; g^\parallel_{|\ell\pm 1|,\ell},
  \; g^\tau_{\ell,\ell\neq 0,0} .
\end{equation}
The plots in \fig{fig:g_functions} show some of them.
Of particular interest are the limiting behaviors, already explored in
\rcite{DPRD20}. When $\lambda\to 0$, the
functions~(\ref{eq:def_littlec_ell}) behave exponentially,
\begin{equation}
  \label{eq:cell_smallxi}
  c_\ell (s,\lambda) \approx -\frac{\lambda^3}{s} \mathrm{e}^{(1-s)/\lambda} .
\end{equation}
Therefore, the integrals in the definitions~(\ref{eq:gfunc}) of the
$g$--functions are exponentially dominated by the lower limit ($s=1$)
and can be thus approximated using Watson's lemma (see, e.g.,
\cite{BeOr78}) in order to obtain the leading asymptotic behavior,
\begin{equation}
  g^{\perp}_{\ell \ell'} \approx 3 \lambda^6 ,
  \quad
  g^{\parallel}_{\ell \ell'} \approx \frac{3}{2} \lambda^5 ,
  \quad
  g^{\tau}_{\ell \ell'} \approx 3 \lambda^5 ,
\end{equation}
(this is \eq{eq:g_small_xi}), which turns out to be independent of the
value of the indices $\ell, \ell'$.

In the opposite limit, $\lambda\to \infty$, one has to address separately
the contributions associated to $V^{(\mathbb{A})}$ and
$V^{(\mathbb{S})}$, i.e., following from the
decomposition~(\ref{eq:QAS}). One first notices that the
functions~(\ref{eq:def_littlec_ell}) behave in this limit as
\begin{equation}
  \label{eq:cell_largexi}
  c_\ell (s,\lambda) \approx -\frac{\lambda^2}{(\ell +1) s^{\ell+1}} ,
\end{equation}
and all the integrals that comply with the
constraints~(\ref{eq:gselec}) are convergent. Thus, all the
$g$--functions scale as $\lambda^2$ in this limit. Since the velocity
scales $V^{(\mathbb{S})}$ and $\Omega^{(\mathbb{S})}$ are always
accompanied by a factor $\lambda^{-2}=(R/\lambda_D)^2$ in
expressions~(\ref{eq:phoreticVel}), this gives a $\lambda$--independent
value for the phoretic velocity stemming from the
surface-charge--driven mechanism.

One would be tempted to conclude that the velocity due to the
correlation--driven mechanism will therefore grow as $\lambda^2$. This is
however not the case because the leading asymptotic contributions
cancel altogether when inserted in the
expressions~(\ref{eq:phoreticVel}). This can be easily understood as
follows: if the limit $\lambda_D = R\lambda \to\infty$ is taken in
\eq{eq:Helm_QA}, the boundary--value problem for the field
$Q^{(\mathbb{A})} (\br)$ becomes the same as the
problem~(\ref{eq:bvp_M}) for the field $M(\br)$, and one can write
$Q^{(\mathbb{A})} = M/\beta c_0$, so that the contribution to the
force field~(\ref{eq:force2}) is a perfect gradient, and thus
irrelevant. This means that this contribution to the phoretic velocity
is determined by the next-to-leading asymptotic behavior of the $g$
functions. This was done explicitly in Sec.~VI of the Supplementary
Material of \rcite{DPRD20} for the functions $g^\perp, g^\parallel$;
the same method can be applied straightforwardly to $g^\tau$. The final
conclusion is that also this piece of the velocity adquires a finite,
$\lambda$--independent value in this limit.

%
%